\documentclass[a4paper,12pt]{article}

\usepackage{amsmath}
\usepackage{amsfonts}
\usepackage{amssymb}

\begin{document}

\begin{center}
{\large\bf Inhomogeneous modes in the PT-symmetric quantum cosmology}

{O.O. Novikov$^{1,\P}$}

$^1${St. Petersburg State University, 7/9 Universitetskaya nab., 199034, St. Petersburg, Russia}

$^\P${E-mail: o.novikov@spbu.ru}
\end{center}

\centerline{\bf Abstract}
We consider the possibility to solve the issues of the phantom field cosmology by means of the PT-symmetric quantum theory. The Born-Oppenheimer approximation is applied to the Wheeler-DeWitt equation to study the inhomogeneous fluctuations over the homogeneous minisuperspace. The evolution of the longitudinal inhomogeneous modes in WKB-time is described using a time-dependent pseudo-Hermitian effective Hamiltonian.

\section{Introduction}

The nature of the Dark Energy remains to be one of the most interesting issues of the modern cosmology. One of the possibilities is that it is produced by so-called phantom matter with $w=p/\epsilon$ going below $-1$ at least at some stage of the cosmological evolution. E.g. one may consider a two-field model with one of the scalar field $\xi$ having wrong-sign kinetic term\cite{quintom},
\begin{equation}
L=\frac{1}{2}\partial_\mu\phi\partial^\mu\phi-\frac{1}{2}\partial_\mu\xi\partial^\mu\xi-V(\phi,\xi),\label{lagphant}
\end{equation}
However as the energy is not bounded from below this leads to severe instabilities and the cosmological evolution in such models generally ends with unlimited expansion in the finite cosmic time (so-called Big Rip).

It is curious that some of the seemingly unstable and even complex classical Hamiltonians with $PT$-symmetry can still possess purely real positive energy spectrum \cite{bender}. The most well-known example is,
\begin{equation}
\hat{H}=p^2+x^2(ix)^\epsilon,\quad \hat{H}^{PT}=\hat{H}
\end{equation}
It happens that for certain non-standard norm $(\psi,\tilde{\psi})$ such Hamiltonian generate a unitary evolution. Such $PT$-symmetric Hamiltonians are the particular case of the so-called pseudo-Hermitian operators related to the Hermitian ones by non-unitary similarity operator $\eta$,
\begin{equation}
\hat{H}=\eta^{-1}\hat{h}\eta,\quad \hat{h}=\hat{h}^\dagger,\quad (\psi,\tilde{\psi})=\langle\psi|\eta^\dagger\eta|\tilde{\psi}\rangle\label{Dyson}
\end{equation}
However the Hermitian form usually happens to be very complicated and it is much easier to study such models in the non-Hermitian form.

It was proposed in \cite{AAA1,AAA2} to apply this idea to the cosmology by replacing (\ref{lagphant}) with a classically equivalent model with the field $\tilde{\phi}=-i\xi$.
\begin{equation}
L=\frac{1}{2}\partial_\mu\phi\partial^\mu\phi+\frac{1}{2}\partial_\mu\tilde{\phi}\partial^\mu\tilde{\phi}-V(\phi,i\tilde{\phi}),
\end{equation}
Then the kinetic energy is positive but the potential is complex. To possess real energy density and pressure the classical solution for $\tilde\phi$ should be purely imaginary. In contrast the perturbations are considered along the real axis, $\tilde{\phi}=i\xi_{class}+\delta\phi$. Such perturbations near classical trajectory happen to possess positively definite effective Hamiltonian. To separate such fields from the usual phantoms we coin a new name - PTom. A realization of this idea as a fully quantum model restricted to the homogeneous minisuperspace was considered in \cite{ourspringer}. In this short paper we elucidate how this approach may work for the longitudinal inhomogeneous modes.

\section{The model of the quintessence and phantom field}

In order to fit observations we need a composition of two scalar fields: quintessence and PTom ones. Let's consider the following model,\cite{ourspringer}
\begin{eqnarray}
S&=&\int d^4x\,\sqrt{-g}\biggl(-\frac{1}{2\kappa^2}R
+\frac{1}{2}M_{\Phi\Phi}\partial_\mu\Phi\partial^\mu\Phi
+\frac{1}{2}M_{\tilde{\Phi}\tilde{\Phi}}\partial_\mu\tilde{\Phi}\partial^\mu\tilde{\Phi}\nonumber\\
&+&iM_{\Phi\tilde{\Phi}}\partial_\mu\Phi\partial^\mu\tilde{\Phi}-V(\Phi,\tilde{\Phi})\biggr),
\label{action}
\end{eqnarray}
where $\kappa^2=8\pi G=M_{Pl}^{-2}$. We take all parameters to be real and assume that $\Big(V(\Phi,-\tilde{\Phi})\Big)^\ast=V(\Phi,\tilde{\Phi})$ to preserve the following symmetry,
\begin{equation}
\mathcal{PT}:\quad t\mapsto-t,\quad i\mapsto-i,\quad \Phi\mapsto\Phi,\quad \tilde{\Phi}\mapsto-\tilde{\Phi} .
\end{equation}

At the quadratic order the transverse (3d tensor and vector) modes decouple from the longitudinal (3d scalar) ones. In this paper we will restrict ourselves to only the longitudinal inhomogeneous sector. We consider the following metric and field ansatz,
\begin{equation}
ds^2=(N^2(t)+s(t,x))dt^2+2(\partial_k v(t,x))dt dx^k-e^{2\rho}\Big(\delta_{ij}+h(t,x)\delta_{ij}+\partial_i\partial_jE(t,x)\Big)dx^i dx^j
\end{equation}
\begin{equation}
\Phi(t,\vec{x})=\Phi(t)+\phi(t,\vec{x}),\quad \tilde{\Phi}=\tilde{\Phi}(t)+\tilde{\phi}(t,\vec{x})
\end{equation}
The variables $\{Q_A\}\equiv\{N,\rho,\Phi,\tilde{\Phi}\}$ are the coordinates on the restricted minisuperspace whereas $\{\chi_a\}\equiv\{s,v,h,E,\phi,\tilde{\phi}\}$ are treated as small fluctuations. We will take the following partial gauge fix,
\begin{equation}
h=E=0
\end{equation}
To proceed we perform IR regulazation to the finite volume $\mathcal{V}$. Let us decompose the remaing fluctuation variables into the superposition of the orthonormal eigenfunctions of the Laplace operator acting on this regularized space.
\begin{equation}
\chi_a(t,\vec{x})=\sum_{n} \chi_{a}^{(n)}(t) f_n(\vec{x}),\quad -\Delta f_n(\vec{x})=\Omega_n^2 f_n(\vec{x}),\quad \int\limits_{\mathcal{V}} d^3x f_n(\vec{x})f_m(\vec{x})=\delta_{nm}.
\end{equation}

The variables $N$,$v$ and $s$ are not dynamical and generate constraints. As usual in general relativity one of the constraints is the Hamiltonian that up to the quadratic order in fluctuations takes the form,
\begin{equation}
H=\mathcal{V}H_0+\sum_n H_2^{(n)}\simeq 0
\end{equation}
\begin{equation}
H_0=-\frac{\kappa^2}{12\mathcal{V}^2}p_\rho^2+\frac{M_{\tilde{\Phi}\tilde{\Phi}}}{2\mathcal{V}^2\mathcal{D}}p_{\Phi}^2+\frac{M_{\Phi\Phi}}{2\mathcal{V}^2\mathcal{D}}\tilde{p}_{\Phi}^2-i\frac{M_{\Phi\tilde{\Phi}}}{\mathcal{V}^2\mathcal{D}}p_{\Phi}\tilde{p}_{\Phi}+V(\Phi,\tilde{\Phi}),\quad \mathcal{D}=M_{\Phi\Phi}M_{\tilde{\Phi}\tilde{\Phi}}+M_{\Phi\tilde{\Phi}}^2
\end{equation}
\begin{align}
H_2=&\frac{M_{\tilde{\Phi}\tilde{\Phi}}}{2\mathcal{V}\mathcal{D}}(p_{\phi}^{(n)})^2+\frac{M_{\Phi\Phi}}{2\mathcal{V}\mathcal{D}}(\tilde{p}_{\phi}^{(n)})^2-i\frac{M_{\Phi\tilde{\Phi}}}{\mathcal{V}\mathcal{D}}p_{\phi}^{(n)}\tilde{p}_{\phi}^{(n)}+\frac{\kappa^2}{48\mathcal{V}^2}s^2p_\rho^2\nonumber\\
&+\frac{s}{2}\Bigg[\frac{M_{\tilde{\Phi}\tilde{\Phi}}}{\mathcal{V}\mathcal{D}}p_{\Phi}p_{\phi}^{(n)}+\frac{M_{\Phi\Phi}}{\mathcal{V}\mathcal{D}}\tilde{p}_{\Phi}\tilde{p}_{\phi}^{(n)}-i\frac{M_{\Phi\tilde{\Phi}}}{\mathcal{V}\mathcal{D}}(p_{\Phi}\tilde{p}_{\phi}+p_{\phi}\tilde{p}_{\Phi})+e^{6\rho}\frac{\partial V}{\partial\Phi}\phi^{(n)}++e^{6\rho}\frac{\partial V}{\partial\tilde{\Phi}}\tilde{\phi}^{(n)}\Bigg]\nonumber\\
&+\frac{e^{6\rho}}{2}\frac{\partial^2V}{\partial\Phi^2}(\phi^{(n)})^2+\frac{e^{6\rho}}{2}\frac{\partial^2V}{\partial\tilde{\Phi}^2}(\tilde{\phi}^{(n)})^2+e^{6\rho}\frac{\partial^2V}{\partial\Phi\partial\tilde{\Phi}}\phi^{(n)}\tilde{\phi}^{(n)}
\end{align}
In addition the following constraints should hold for each inhomogeneous mode,
\begin{align}
\mathcal{H}_s^{(n)}=&\frac{\kappa^2}{12\mathcal{V}^2}sp_\rho^2+\frac{M_{\tilde{\Phi}\tilde{\Phi}}}{\mathcal{V}\mathcal{D}}p_{\Phi}p_{\phi}^{(n)}+\frac{M_{\Phi\Phi}}{\mathcal{V}\mathcal{D}}\tilde{p}_{\Phi}\tilde{p}_{\phi}^{(n)}-i\frac{M_{\Phi\tilde{\Phi}}}{\mathcal{V}\mathcal{D}}(p_{\Phi}\tilde{p}_{\phi}+p_{\phi}\tilde{p}_{\Phi})\nonumber\\
&+e^{6\rho}\frac{\partial V}{\partial\Phi}\phi^{(n)}++e^{6\rho}\frac{\partial V}{\partial\tilde{\Phi}}\tilde{\phi}^{(n)}\simeq 0,
\end{align}
\begin{equation}
\mathcal{H}_v^{(n)}=\frac{s}{6}p_\rho+\frac{M_{\tilde{\Phi}\tilde{\Phi}}}{\mathcal{D}}p_{\Phi}\phi^{(n)}+\frac{M_{\Phi\Phi}}{\mathcal{D}}\tilde{p}_{\Phi}\tilde{\phi}^{(n)}-i\frac{M_{\Phi\tilde{\Phi}}}{\mathcal{D}}(p_{\Phi}\tilde{\phi}+\phi\tilde{p}_{\Phi})\simeq 0
\end{equation}
The costraints $\mathcal{H}_{s}^{(n)}$ and $\mathcal{H}_{v}^{(n)}$ correspond to the inhomogeneous Hamiltonian and momentum constraints of the ADM formalism correspondingly.

\section{The quantum theory of the fluctuations}

We now perform the Dirac quantization of the constraint system obtained above. We derive the system of the Wheeler-DeWitt (WdW) equations in the following form,
\begin{align}
\Big(\mathcal{V}\hat{H}_0+\sum_n\hat{H}_2^{(n)}\Big)\Psi(\{Q_A\},\{\chi_a\}\Big)=0,\\
\hat{\mathcal{H}}^{(n)}\Psi(\{Q_A\},\{\chi_a\}\Big)=0,\quad \hat{\tilde{\mathcal{H}}}^{(n)}\Psi(\{Q_A\},\{\chi_a\}\Big)=0
\end{align}
Then we use the Born-Oppenheimer approximation \cite{kiefer}-\cite{kamenshik} and separation of the fluctuation variables corresponding to the different inhomogeneous modes,
\begin{equation}
\Psi(\{Q_A\},\{\chi_a\}\Big)\simeq \Psi_0(\{Q_A\})\prod_n\Psi_2^{(n)}(\{Q_A\},\{\chi_a^{(n)}\})
\end{equation}
Decomposing the WdW equations into different orders of $\kappa^3/\mathcal{V}$ we first obtain the WdW equation in the restricted minisuperspace,
\begin{equation}
\hat{H}_0\Psi_0=\left(\frac{\kappa^2}{12\mathcal{V}^2}\partial_\rho^2-\frac{M_{\tilde{\Phi}\tilde{\Phi}}}{2\mathcal{V}^2\mathcal{D}}\partial_{\Phi}^2-\frac{M_{\Phi\Phi}}{2\mathcal{V}^2\mathcal{D}}\partial_{\tilde{\Phi}}^2+i\frac{M_{\Phi\tilde{\Phi}}}{\mathcal{V}^2\mathcal{D}}\partial_{\Phi}\partial_{\tilde{\Phi}}+V(\Phi,\tilde{\Phi})\right)\Psi_0=0,
\end{equation}
Please note that the large volume limit $\mathcal{V}/\kappa^3\rightarrow\infty$ serves as an quasiclassical limit. Assuming that we consider the quasiclassical wavepacket far from the turning point we may use the WKB approximation for the minisuperspace part of the wavefunction,
\begin{equation}
\Psi_0(\{Q_A\})=\psi_0(\{Q_A\}) \exp\Big(i\mathcal{V}S(\{Q_A\})\Big)
\end{equation}

We will neglect backreaction assuming that $\hat{H}_0\Psi_2^{(n)}=0$. Then for the fluctuation part of the wavefunction we obtain,
\begin{equation}
\hat{\mathcal{H}}^{(n)}\Psi_2^{(n)}=0,\quad \hat{\tilde{\mathcal{H}}}^{(n)}\Psi_2^{(n)}=0,\quad
\frac{i}{\mathcal{V}}\partial_\tau\Psi_2^{(n)}=\hat{H}_2^{(n)}\Psi_2^{(n)},\label{WKBSchrodinger}
\end{equation}
where $\partial_\tau$ is the derivative in the so-called WKB-time that is directed along the classical trajectory\cite{kiefer},
\begin{align}
\frac{1}{\mathcal{V}}\partial_\tau=G^{AB}(\partial_A S)\partial_B,
\end{align}
with $\mathcal{G}^{AB}$ being the metric on the minisuperspace $\{Q_A\}$,
\begin{equation}
\mathcal{G}^{AB}X_A Y_B\equiv \frac{\kappa^2}{12}\partial_\rho^2-\frac{M_{\tilde{\Phi}\tilde{\Phi}}}{2\mathcal{D}}X_{\Phi}Y_{\Phi}-\frac{M_{\Phi\Phi}}{2\mathcal{D}}X_{\tilde{\Phi}}Y_{\tilde{\Phi}}+i\frac{M_{\Phi\tilde{\Phi}}}{2\mathcal{D}}(X_{\Phi}Y_{\tilde{\Phi}}+X_{\tilde{\Phi}}Y_{\Phi}),
\end{equation}

The quadratic term $H_2^{(n)}$ may be then interpreted as the effective Hamiltonian generating evolution of the fluctuation part of the wavefunction along the family of the classical solutions defined by WKB-action $S$. We can then try to treat it as the time-dependent pseudo-Hermitian Hamiltonian using the generalization of \eqref{Dyson} following \cite{TDDyson},
\begin{equation}
\hat{H}_2^{(n)}=\Big(\eta_n(\{Q_A\})\Big)^{-1} \hat{h}^{(n)}(\{Q_A\})\eta_n(\{Q_A\})-i\Big(\eta_n(\{Q_A\})\Big)^{-1}\partial_\tau\Big(\eta_n(\{Q_A\})\Big),\label{TDPH}
\end{equation}
where both $\hat{h}^{(n)}$ and $\eta_n$ are Hermitian operators. Note that because of the second term in the time-dependent case the resulting non-Hermitian Hamiltonian may actually not be observable and may have complex eigenvalues but still describe some unitary evolution with respect to the norm $\langle\psi|\eta^\dagger\eta|\phi\rangle$.

Let us consider the specific potential and kinetic matrix that allows exact classical and quantum homogeneous solutions \cite{ourspringer,ourtmph},
\begin{equation}
V(\Phi,\tilde{\Phi})=Ve^{\Phi}-\tilde{V}e^{i\tilde{\Phi}},\, \frac{M_{\Phi\tilde{\Phi}}}{\mathcal{D}}
=6\kappa^2,\, D=\frac{M_{\tilde{\Phi}\tilde{\Phi}}}{\mathcal{D}}-6\kappa^2,\, \tilde{D}=\frac{M_{\Phi\Phi}}{\mathcal{D}}+6\kappa^2
\end{equation}
In the weak gravity limit $\kappa\rightarrow 0$ after solving the constraint $\mathcal{H}_s^{(n)}$ the effective Hamiltonian greatly simplifies with $\phi$ and $\tilde{\phi}$ being decoupled,
\begin{equation}
\hat{H}_2^{(n)}=\hat{H}_{\phi,2}^{(n)}+\hat{H}_{\tilde{\phi},2}^{(n)}+O(\kappa),
\end{equation}
\begin{equation}
\hat{H}_{\phi,2}^{(n)}=\frac{D}{2}(p_{\phi}^{(n)})^2+\frac{V}{2}e^{6\rho+\Phi}(\phi^{(n)})^2,\quad
\hat{H}_{\tilde{\phi},2}^{(n)}=\frac{\tilde{D}}{2}(\tilde{p}_{\phi}^{(n)})^2+\frac{\tilde{V}}{2}e^{6\rho+i\tilde{\Phi}}(\tilde{\phi}^{(n)})^2
\end{equation}
Because we consider $\Phi$ and $\phi$ as ordinary Hermitian fields $\hat{H}^{(n)}_{\phi,2}$ is manifestly Hermitian and $\eta_n$ should be nontrivial only on the $\tilde{\phi}$ sector. The remaining part $\hat{H}_{\tilde{\phi},2}$ is a non-Hermitian time-dependent oscillator (or more generally the Swanson Hamiltonian) similar to the problem considered in \cite{Fring,Maamache}. Instead of appealing to their work we suggest somewhat different approach. We start by noting that for purely imaginary solution $\tilde{\Phi}=i\xi_{class}$ this is simply a Hermitian oscillator with time-dependent frequency. Then we decompose $\tilde{\Phi}\simeq i\xi_{class}+\delta\tilde{\Phi}$ and look for the solution in the following form,
\begin{equation}
\hat{h}_{\tilde{\phi}}^{(n)}=\hat{H}_{2,\tilde{\phi}}^{(n)}\Big\vert_{\tilde{\Phi}=i\xi_0},\, \eta_n\equiv e^{X_n}=\exp\Big[\alpha_n(\delta\tilde{\Phi})p_{\tilde{\phi}}^2+\beta_n(\delta\tilde{\Phi})\tilde{\phi}^2+\gamma_n(\delta\tilde{\Phi})(\tilde{p}_{\phi}\tilde{\phi}+\tilde{\phi}\tilde{p}_{\phi})\Big]
\end{equation}
To calculate $\hat{H}_{2,\tilde{\phi}}^{(n)}$ we note that the operators $\{p_{\tilde{\phi}}^2,\tilde{\phi}^2,\tilde{p}_{\phi}\tilde{\phi}+\tilde{\phi}\tilde{p}_{\phi}\}$ form the finite algebra $\mathcal{A}$. Then we can use the following representation of the operator exponentials \cite{Ross},
\begin{equation}
e^{-X_n}\hat{h}_{\tilde{\phi}}^{(n)} e^{X_n}=e^{-\mathrm{ad}_{X_n}}\hat{h}_{\tilde{\phi}}^{(n)},\quad e^{-X_n}\partial_\tau\Big(e^{X_n}\Big)=\frac{1-e^{-\mathrm{ad}_{X_n}}}{\mathrm{ad}_{X_n}}\partial_\tau X_n,
\end{equation}
where the adjoint operator $\mathrm{ad}_X Y\equiv [X,Y]$ may be represented as a finite-dimensional matrix acting on the algebra $\mathcal{A}$.

To simplify the problem let us assume that in terms of the minisuperspace probability measure (which is the question that lays beyound the scope of this paper) $\Psi_0$ is a wavepacket that strongly concentrated near $\tilde{\Phi}=i\xi_{class}$. Then $\delta\tilde{\Phi}$,$\alpha_n$,$\beta_n$ and $\gamma_n$ can be considered a perturbation parameter. From the NLO of (\ref{TDPH}) we obtain the following ODE system,
\begin{equation}
\partial_\tau\alpha_n=2\gamma_n\tilde{D},\quad
\partial_\tau\gamma_n=2\beta\tilde{D}-2\alpha\tilde{V}e^{6\rho-\xi_{class}},\quad
\partial_\tau\beta_n=\frac{\tilde{V}}{2}e^{6\rho-\xi_{class}}\delta\tilde{\Phi}-4\gamma_n\tilde{V}e^{6\rho-\xi_{class}}
\end{equation}
The solutions can be found numerically using exact classical solutions $\xi_{class}$ found in \cite{ourspringer}.

The norm of the fluctuation part conserved in WKB time is constructed as,
\begin{equation}
\Big(\Psi_2|\tilde{\Psi}_2\Big)_{\{Q_A\}}=\langle\Psi_2|\eta^\dagger\eta|\tilde{\Psi}_2\rangle_{\{Q_A\}}
\end{equation}
\begin{equation}
\langle\Psi_2|\tilde{\Psi}_2\rangle_{\{Q_A\}}=\prod_n\left(\int\limits_{\mathbb{R}^2} d\phi_n d\tilde{\phi}_n\right)\Psi_2^\ast(\{Q_A\},\{\chi_a^{(n)}\})\tilde{\Psi}_2(\{Q_A\},\{\chi_a^{(n)}\})
\end{equation}

The wide wavepacket case and implications of this result for the construction of the full probability measure will be considered in the future work.

\section*{Acknowledgment}
The work is done with financial support by grant RFBR, project 16-02-00348. The author is grateful to A.A.Andrianov, Chen Lan and A.V.Golovnev for helpful discussions.


\begin{thebibliography}{99}
\bibitem{quintom}
  {\it Cai Y.F., Saridakis E.N., Setare M.R. and Xia J.Q.,} Quintom Cosmology: Theoretical implications and observations // Phys.\ Rept.\ 2010. V. 493. P. 1.
\bibitem{bender}
  {\it Bender C.M.} Making sense of non-Hermitian Hamiltonians
// Rept. Prog. Phys. 2007. V. 70. P. 947.
\bibitem{AAA1}
  {\it Andrianov A.A., Cannata F. and Kamenshchik A.Y.,}
  Phantom universe from CPT symmetric QFT,
  Int. J. Mod. Phys. D 2006. V. 15. P. 1299.
\bibitem{AAA2}
  {\it Andrianov A.A., Cannata F. and Kamenshchik A.Y.,}
  Complex Lagrangians and phantom cosmology,
  J. Phys. A. 2006 V. 39. P. 9975.
\bibitem{ourspringer}
  {\it Andrianov A.A., Lan C., Novikov O.O.,}
  PT Symmetric Classical and Quantum Cosmology,
  In: Bagarello F., Passante R., Trapani C. (eds) Non-Hermitian Hamiltonians in Quantum Physics. Springer Proceedings in Physics. V. 184. Springer, Cham. 2016.
\bibitem{kiefer}
  {\it Kiefer C.,}
  Quantum gravity,
  Oxford Uni. Press, II edition, 2007. P. 168.
\bibitem{BO1}
 {\it Brout R.},
  On the Concept of Time and the Origin of the Cosmological Temperature,
  Found.\ Phys.\  1987. V. 17. P. 603.
\bibitem{BO2}
 {\it Brout R. and Venturi G.},
  Time in Semiclassical Gravity,
  Phys.\ Rev.\ D 1989. V. 39. P. 2436.
\bibitem{BO3}
  {\it Kamenshchik A.Y., Tronconi A. and Venturi G.},
  Inflation and Quantum Gravity in a Born-Oppenheimer Context,
  Phys.\ Lett.\ B. 2013. V. 726 P. 518
\bibitem{BO4}
  {\it Kamenshchik A.Y., Tronconi A. and Venturi G.},
  Signatures of quantum gravity in a Born–Oppenheimer context,
  Phys.\ Lett.\ B. 2014. V. 734. P. 72.
\bibitem{BO5}
  {\it Kiefer C. and Kraemer M.},
  Quantum Gravitational Contributions to the CMB Anisotropy Spectrum,
  Phys.\ Rev.\ Lett.\ 2012. V. 108. P. 021301.
\bibitem{BO6}
  {\it Brizuela D., Kiefer C. and Kraemer M.},
  Quantum-gravitational effects on gauge-invariant scalar and tensor perturbations during inflation: The slow-roll approximation,
  Phys.\ Rev.\ D. 2016. V. 94. P. 123527.
\bibitem{kamenshik}
 {\it Kamenshchik A.Y., Tronconi A. and Venturi G.},
  The Born-Oppenheimer Method, Quantum Gravity and Matter,
  [arXiv:1709.10361].
\bibitem{TDDyson}
  {\it Fring A. and Moussa M.H.Y.,}
  Unitary quantum evolution for time-dependent quasi-Hermitian systems with nonobservable Hamiltonians
// Phys. Rev. A. 2016. V. 93. P. 042114.
\bibitem{ourtmph}
  {\it Andrianov A.A., Novikov O.O., Lan C.,}
  Quantum cosmology of the multi-field scalar matter: some exact solutions
  // Theor. Math. Phys. 2015. V. 184. P. 1224.
\bibitem{Fring}
  {\it Fring A. and Moussa M.H.Y.,}
   The non-Hermitian Swanson model with a time-dependent metric
//  	Phys. Rev. A. 2016. V. 94. P. 042128.
\bibitem{Maamache}
  {\it Maamache M., Djeghiour O.K., Mana N., Koussa W.,}
  Quantum Evolution of the Time-Dependent Non-Hermitian Hamiltonians: Real Phases,
  [arXiv: 1705.06341].
\bibitem{Ross}
{\it Rossmann, W.},
 Lie Groups – An Introduction Through Linear Groups,
Oxford Uni. Press, 2002.
   
\end{thebibliography}
\end{document}